\documentstyle[prl,aps,multicol,epsf]{revtex}
\twocolumn
\begin{document}
\draft
\title{Universal quantum gates based on a pair of orthogonal cyclic states:
Application to NMR systems}
\author{Shi-Liang Zhu$^{1,2}$\thanks{Email address: szhu@hkucc.hku.hk}
and 
Z. D. Wang$^{1,3}$ \thanks{Email address: zwang@hkucc.hku.hk}
}
\address{ 
$^{1}$Department of Physics, University of Hong Kong, Pokfulam Road,
Hong Kong, China\\
$^{2}$Department of Physics, South China Normal University,
Guangzhou, China\\
$^{3}$ Department of Material Science and Engineering, University of 
Science and Technology of China, Hefei, China
}
\address{\mbox{}}
\address{\parbox{14cm}{\rm \mbox{}\mbox{}
We propose an experimentally feasible scheme to
achieve quantum computation based on a
pair of orthogonal cyclic states.
In this scheme, quantum gates can be implemented based on the total
phase accumulated in cyclic evolutions. In particular,
geometric quantum computation may be achieved by eliminating
the dynamic phase accumulated in the whole evolution. Therefore,
both
dynamic and geometric operations for quantum computation 
are workable in the present theory.
Physical implementation of this set of gates is designed for
NMR systems. Also interestingly, we show that
a set of universal geometric quantum gates
in NMR systems may be realized in one cycle by simply choosing
specific parameters of the external rotating magnetic fields.
In addition, we demonstrate explicitly a multi-loop method to remove
the dynamic phase in geometric quantum gates.
}}
\address{\mbox{}}
\address{\parbox{14cm}{\rm PACS numbers:
03.67.Lx, 03.65.Vf, 03.67.Pp}}
\maketitle

\newpage
\narrowtext

\section{Introduction}

Building a practical quantum computer with a large number
of qubits has recently attracted much attention.
For  realization of a universal quantum computer,
there are certain minimum requirements:
the storage of quantum information
in a set of two-level systems--qubits, the processing of
this information using quantum gates, and a
mean of final readout\cite{DiVincenzo}.
So far, a number of systems has been proposed as potentially
viable quantum computer models,  including
trapped ions\cite{Cirac}, 
cavity quantum electrodynamics\cite{Pellizzari},
nuclear magnetic resonce(NMR)\cite{Gershenfeld},
and low-capacitance Josephson
juctions\cite{Shnirman,Makhlin,Nakamura,Falci},
{\it etc}..

An essential requirement
in quantum computation is to maintain quantum coherence
in a computing system,
since the coherent interference pattern
between the multitude of superpositions
is necessary for taking advantage of quantum parallelism.
However, the coupling of
a quantum system to its environment leads to
the so-called decoherence process,
in which encoded quantum information is lost
to the environment.
The error rates of the individual gate operations
should be less than
$10^{-4}$ to assure that the quantum computer
works fault-tolerantly\cite{Preskill}.
To accomplish the required precision, the decoherence time
of the system has to be much longer than
the operation time required for computing.
How to suppress the infamous decoherence effects is one main task
for quantum computing.

One of schemes to correct the errors caused by decoherence
is quantum error-correcting
codes\cite{Shor95,Bennett96,Steane96},
through which originally
encoded information can be recovered by
suitable encondings and measurements of qubits.
An alternative approach to avoid decoherence
has been proposed in Refs. \cite{Duan_prl1997,Zanardi1997},
where decoherence-free states have been used as qubits.
The decoherence-free space is
a subspace
which is inherently immune to unwanted noise.
In addition, evolution of the system must not drive
the state out of the decoherence-free
space. So far, all of these strategies require
extra-physical resources, such as additional manipulations or
encoding one logical qubit by several
ancillary physical qubits.

Another attractive strategy for 
fault-tolerant quantum computation is based on a topological
idea\cite{Kitaev}, where gate operations depend only
on global features of the control process, and are therefore
largely insensitive to local inaccuracies and fluctuations.
A significant advance in this direction is made by the
so-called geometric quantum computation\cite{Zanardi1999}.
In this kind of scheme, a universal set of quantum gates
may be realized by pure geometric phases,
which depend only on the geometry of the path 
executed\cite{Berry,Aharonov,Li},
and therefore provides a possibility to perform quantum gate
operations by an intrinsically
fault-tolerant way\cite{Zanardi1999,Jones,Duan}.

Several basic ideas of adiabatic geometric
quantum computation by using NMR\cite{Jones}, 
superconducting nanocircuits\cite{Falci},
trapped ions\cite{Duan}, or semiconductor
nanostructures\cite{Solinas} were proposed.
However, the adiabatic evolution appears to be quite special,
and thus the nonadiabatic correction on  the phase shift
may need to be considered in some realistic systems
as it may play a significant role in a whole process~ 
\cite{Aharonov,Zhu_prl2000}.
A serious disadvantange of the adiabatic
approach is that the evolution time must be
much longer than the typical operation time
$\tau_0$ of the qubit system,
while the evolution must be completed within the
decoherence time,
which leads to an intrinsical time limitation
on the operation of quantum gate.
Therefore, a generalization to nonadiabatic cases
is  valuable and important in controlling quantum gates.

Recently, nonadiabatic geometric quantum computation
has been proposed theoretically\cite{Wang,Zhu_prl2002},
and detection of the conditional nonadiabatic geometric phase
shifts for quantum gates using NMR
is also experimentally reported\cite{Du}. Nevertheless a
systematic study of this topic, especially the appication to NMR systems, 
is still highly desirable.
In this paper, we propose an experimentally feasible nonadiabatic scheme to
achieve a universal set of quantum gates \cite{Lloyd} based on a
pair of orthogonal cyclic states.
In this scheme, quantum gates may be implemented based on either 
the total phase accumulated in the cyclic evolution 
or the geometric Aharonov-Anandan (AA) phase shift \cite{Aharonov}
after eliminating the dynamic phase. Therefore,
quantum computation implemented by
dynamic and geometric operations can be unified in the present
theory. In addition, 
physical implementation of this set of gates is designed
in detail  for NMR systems, in which 
the qubits considered are spin-$1/2$ particles
in the presence of a magnetic field rotating uniformly
around a fixed axis. Although the Schr\"{o}dinger equation
of this system was solved exactly long time ago\cite{Rabi},
and the AA phase was previously obtained explicitly in
Refs.\cite{Zhu_prl2000,Lin},
we here derive all phase shifts 
explicitly and show that
they may be
applicable in achieving a universal set of logical gates.
Moreover, the nonadiabatic
geometric computation may be experimentally
achieved just by simply choosing
specific controllable parameters, with
the cyclic states as a pair of dark states\cite{Wang}.

The paper is organized as follows. In Sec. II,
we discuss general aspects of the geometric phase and cyclic
evolution, and then present a theory
applicable for achieving a universal set of quantum gates
based on a pair of orthogonal cyclic states.
In Sec. III, the theory is applied to 
a viable NMR quantum computer.
The paper ends with a brief summary.

\section{Implementation of quantum gates with a pair of orthogonal
cyclic states}

For universal quantum computation, we need to achieve two
kinds of noncommutable single-qubit gates and one nontrivial
two-qubit gate \cite{Lloyd}. Thus we here consider
only two-qubit systems.
A general Hamiltonian for two
qubits may be expressed as
\begin{equation}
\label{H_two_bit}
\hat{H}=-\frac{1}{2}\mu\sum\limits_{i=1}^3[\sigma_i^{(1)} B_i^{(1)}(t)
+\sigma_i^{(2)} B_i^{(2)}(t)+J_i(t) \sigma_i^{(1)} \sigma_i^{(2)}],
\end{equation}
where $\sigma_i^{(k)}\ (k=1,2)$ are the Pauli operators for qubit
$k$, $B_i^{(k)} (t)$ are local (real or fictitious)
magnetic fields acting on $k$-th qubit,
and $J_i$ represents the strength of the interaction
between two qubits.

\subsection{Cyclic evolution and geometric phases}

Before the design of 
quantum gates,
we present first general aspects of the
cyclic evolution in a qubit system.
A cyclic evolution is referred to as that the state of the system
returns to its original state after evolution.
Mathematically, a normalized state $|\psi (t) \rangle$ is cyclic
in the interval $[0,\tau]$ if and only if
$$
|\psi(\tau)\rangle=e^{i\gamma} | \psi(0) \rangle,
$$ 
with $\gamma$ being a real number. 
The total phase $\gamma$ acquired in the evolution would contain
both geometric and  dynamic components, denoted as
$\gamma_g$ and $\gamma_d$, respectively.
Usually, not all states take cyclic evolutions.
A sufficient but not necessary
condition for cyclic evolution is that
the initial state is a nondegenerate energy eigenstate of
a cyclic Hamiltonian which changes
adiabatically.

We first focus on the cyclic evolution in a qubit system.
At this stage, only one term
\begin{equation}
\label{H_sq}
\hat{H}=-\mu{\bf B}\cdot\stackrel{\rightarrow}{\sigma}/2
\end{equation}
needs to be considered.
Here ${\bf B}$ denotes
the total magnetic field felt by the qubit, which
may include a real external field as well as an
effective magnetic field
induced by the interaction between
different qubits.
The Hamiltonian $\hat{H}$ is chosen to go through a cyclic evolution
with period $\tau$ in the parameter space $\{ {\bf B} \}$.

We here present how to calculate all kinds of phase shifts.
The dynamic phase may be directly calculated from its definition
given by
\begin{equation}
\label{dynamic}
\gamma_d=-\frac{1}{\hbar}\int_0^\tau 
\langle \psi (t)| \hat{H} |\psi(t)\rangle dt,
\end{equation}
while the geometric part is not easy to derive.
We first address a method to calculate
adiabatic Berry's phase.
By adiabatically changing $\hat{H}$
around a circuit in $\{ {\bf B} \}$,
the eigenstate will accumulate an adiabatic Berry's phase
$\gamma_B=\mp \Omega_s/2$, where the
signs $\mp$ depend on whether the system is in
the eigenstate aligned with or against the
field, and $\Omega_s$ is the solid angle subtended by
the magnetic field
at the degeneracy ${\bf B}=0$\cite{Berry}.
$\Omega_s$ can be derived as
\begin{equation}
\label{solid}
\Omega_s=\int_0^\tau \frac{B_x\partial_t B_y-B_y\partial_t B_x}
{|{\bf B}|(B_z+|{\bf B}|)},
\end{equation}
under the condition of a closed trajectory with
${\bf B}(\tau)={\bf B}(0)$\cite{Zhu_prl2002}.

Since the requirement of the adiabatic
evolution could be stringent, a generalization to nonadiabatic
case is more desirable.
The generalization of adiabatic Berry's phase to a nonadiabatic
cyclic evolution was introduced 
in Ref. \cite{Aharonov}, where
a general geometric phase
$\gamma_{g}=\gamma-\gamma_d$ is defined
as
\begin{equation}
\label{AA_phase}
\gamma_{g}=i\int_0^\tau \langle\tilde{\psi} (t)|\frac{\partial}{\partial t}|\tilde{\psi} (t)\rangle,
\end{equation}
Here $|\tilde{\psi} (t)\rangle=e^{-if(t)}|\psi(t)\rangle$
with $f(\tau)-f(0)=\gamma$, leading to  
$|\tilde{\psi} (\tau)\rangle=|\tilde{\psi} (0)\rangle$.
The AA phase can be regarded as a geometric phase
associated with a closed curve
in the projective Hilbert space, and
approaches Berry's phase in the adiabatic limit.
The AA formulation applies regardless of the Hamiltonian
$\hat{H}$ being cyclic or adiabatic; it
depends only on the cyclic evolution of the system itself.

Normally, Eq. (\ref{AA_phase})
is not directly used to calculate the geometric phase
accumulated in a cyclic evolution.
We here present an alternative method to calculate the
non-adiabatic geometric phases. This approach is more convenient
for qubit systems discussed in this paper. 
For a spin-$1/2$ particle in the presence of an
arbitrary magnetic field, the nonadiabatic  cyclic
AA phase is just the solid angle
determined by the evolution curve in the projective Hilbert
space--a unit sphere $S^2$.
Any two-component 'spin' state
$|\psi\rangle=
[e^{-i\varphi/2}\cos(\theta/2),\ e^{i\varphi/2}\sin(\theta/2)]^T$
may be mapped into
a unit vector
${\bf n}=(\sin\theta \cos\varphi,
\sin\theta \sin\varphi,\cos\theta)$
in the projective Hilbert space
via the relation
${\bf n}=\langle\psi|\stackrel{\rightarrow}{\sigma|}\psi\rangle$,
where $T$ represents the transposition of matrix.
By changing the magnetic field,
the AA phase is given by

\begin{equation}
\label{AA}
\gamma_g=-\frac{1}{2} \int_{C}
(1-\cos\theta)d\varphi,
\end{equation}
where $C$ is along the actual evolution curve on
$S^2$, and is determined by
the equation 

\begin{equation}
\label{path}
\partial_t{\bf n}(t)=
-\mu{\bf B}(t)\times{\bf n}(t)/\hbar.
\end{equation}
This $\gamma_g$ phase recovers Berry's phase in
adiabatic evolution\cite{Zhu_prl2000}.
The cyclic evolution implies that
${\bf n}(t)$ undergoes a closed path in the projective Hilbert space.

We consider a process, in which a pair of orthogonal states
$|\psi_{\pm}\rangle$ can evolve cyclically starting
from $|\psi_{\pm}(0)\rangle$.
A pair of orthogonal state
may be parametrized as 
\begin{equation}
\label{psi_p}
|\psi_{+}\rangle=
{\cos\frac{\theta}{2} e^{-i\varphi/2}
\choose \sin\frac{\theta}{2} e^{i\varphi/2}}
\end{equation}
and
\begin{equation}
\label{psi_m}
|\psi_{-}\rangle=
{-\sin\frac{\theta}{2} e^{-i\varphi/2}
\choose \cos\frac{\theta}{2} e^{i\varphi/2}}.
\end{equation}
Denoting ${\bf n}_{\pm}(t)=\langle\psi_{\pm}(t)
|\stackrel{\rightarrow}{\sigma|}\psi_{\pm}(t)\rangle$,
 it is straightforward
to find that ${\bf n}_{+}(t)=-{\bf n}_{-}(t)$
by using Eqs.(\ref{psi_p}) and (\ref{psi_m}).
For a cyclic evolution,
$|\psi_{\pm}(\tau)\rangle=e^{i\gamma_{\pm}} | \psi_{\pm}(0) \rangle$.
Besides, we have an important relation: $\gamma_{+}=-\gamma_{-}$.
This is because  the dynamic phase
\begin{eqnarray*}
\gamma_{d+} &=& -\frac{1}{\hbar}\int_0^\tau E_{+}(t)dt
=-\frac{1}{\hbar}\int_0^\tau -E_{-}(t)dt\\
&=& -\gamma_{d-}
\end{eqnarray*}
with
$$
E_{\pm}(t)=\langle\psi_{\pm}(t)|H|\psi_{\pm}\rangle
=-\mu{\bf n}_{\pm}(t)\cdot {\bf B}(t),
$$
and the geometric phase $\gamma_g(-{\bf
n}(0))= -\gamma_g({\bf n}(0))$ at any time if the two initial states
correspond to $\pm {\bf n}(0)$~\cite{Zhu_prl2000}.
By taking into account the cyclic condition for
$|\psi_{\pm}\rangle$, we finally have
\begin{equation}
\label{cyclic_U}
U(\tau)|\psi_{\pm}\rangle =\exp(\pm
i\gamma)|\psi_{\pm}\rangle, 
\end{equation}
where $U(\tau)$ is the evolution
operator. Hereafter we denote
$\gamma$,$\gamma_g$, and $\gamma_d$
as the phases for $|\psi_+\rangle$ for brevity.

\subsection{Quantum computation}

We now show how to realize a universal set of quantum gates
based on either the total phases or the  geometric AA phases
accumulated in cyclic evolutions.

\subsubsection{Quantum logical gates}

A quantum logical gate is a unitary operator $U$ acting on the states
of a certain set of qubits, that is, $U$ may be referred to as 
a quantum gate if $|\psi_{out}\rangle=U|\psi_{in}\rangle$
with $|\psi_{in}\rangle$ being the input state and $|\psi_{out}\rangle$
being the output state.
The space of all the possible input and output states makes up the Hilbert
space of states for
the quantum computer. If $\mathcal{H}$ is the
Hilbert space of a single qubit, and $|\psi_i\rangle$  
is a given basis state for the $i$th qubit,
then a basis vector $|\psi\rangle$ for the states
of the quantum register is a tensor product of qubit states
$|\psi\rangle=|\psi_1\rangle \otimes |\psi_2\rangle \otimes \cdots
\otimes |\psi_n\rangle \in {\mathcal{H}}^{ {\otimes} n}$.
$U_N$ is an $N$-qubit gate when $|\psi_{in}\rangle
\in {\mathcal{H}}^{ {\otimes} N}$.
Nevertheless, we need not to implement all
$U_l\ (1\le l \le N)$ but only a universal set of gates ${U_u}$.
A set of gates ${U_u}$ is called universal if any unitary
action $U_l$ can be decomposed into a product of
successive gates in $U_u$. It is shown that
two noncommutable one-qubit (single-qubit) gates
and one nontrivial two-qubit gate consist of a universal set of 
gates\cite{Lloyd}. This universality is
very useful in practice since it
allows us to focus only on how to construct
a universal set of gates.

We first construct the single-qubit gates
by assuming that a pair of orthogonal 
states $|\psi_{\pm}\rangle$ can evolve cyclically.
We write an
arbitrary  input state  as
$|\psi_{in}\rangle=a_{+}|\psi_{+}\rangle+a_{-}|\psi_{-}\rangle$ with
$a_{\pm}=\langle \psi_{\pm}|\psi\rangle$, and express
the two cyclic initial states
as $|\psi_{+}\rangle = \cos \chi/2|0\rangle+
\sin \chi/2|1\rangle$ and
$|\psi_{-}\rangle = -\sin \chi/2|0\rangle+
\cos \chi/2|1\rangle$,
where $|0\rangle$ and $|1\rangle$
constitute the computational basis for
the qubit.
Using Eq.(\ref{cyclic_U}), 
the output state at time $\tau$
is found to be\cite{Zhu_prl2002}
$$|\psi_{out}\rangle=U^{sq}(\chi,\gamma)|\psi_{in}\rangle,$$
where
\begin{equation}
\label{single_U}
U^{sq}(\chi,\gamma)=\left (
\begin{array}{ll}
e^{i\gamma}\cos^2\frac{\chi}{2}+e^{-i\gamma}\sin^2\frac{\chi}{2}
& i\sin\chi \sin\gamma  \\
i \sin\chi \sin\gamma
&  e^{i\gamma}\sin^2\frac{\chi}{2}+e^{-i\gamma}\cos^2\frac{\chi}{2}
\end{array}
\right ).
\end{equation}
For this gate, there exists a relation
$$
[U^{sq}(\chi,\gamma)]^\dagger=
[U^{sq}(\chi,-\gamma)],
$$ 
where the adjoint operation $\dagger$
corresponds to transposition and
complex conjugation of matrix.
Thus the important $\dagger$ operator
for a unitary operation $U$ may be experimentally
achieved by the operation $U$ with the inverse sign
of the phase factor.

It is straightforward to verify that two operations
$U^{(1)}(\chi_1,\gamma_1)$ and $U^{(2)}(\chi_2,\gamma_2)$ are
noncommutable as long as
\begin{equation}
\label{criterion}
\sin\gamma_1\sin\gamma_2\sin(\chi_2-\chi_1)\neq 0. 
\end{equation}
Since two kinds
of noncommutable operations constitute a universal set of
single-qubit gates, we achieve the
universal single-qubit gates by choosing  $\chi_1\not=\chi_2+j\pi$
for any nontrivial phases $\gamma_1$ and $\gamma_2$
($\gamma_{1,2} \not= j\pi$), where $j$ is an integer. For
example, the phase-flip gate
$U_1^{sq}(\gamma_1)
=\exp(-2i\gamma_1|1\rangle\langle1|)$ (up to an irrelevant
overall phase) is accomplished at $\chi=0$; the gate
$U_2^{sq}(\gamma_2)
=\exp(i\gamma_2\sigma_x)$ is obtained at $\chi=\pi/2$, which
produces a spin flip (NOT-operation) when $\gamma_2=\pi/2$ and an
equal-weight superposition of spin states when $\gamma_2=\pi/4$.
$U^{sq}_{1,2}$ are two well-known single-qubit gates.

In terms of the computational
basis $\{|00\rangle$,$|01\rangle$,$|10\rangle$,
$|11\rangle \}$, where the first (second) number represent
the state in the controlled (target) qubit, 
the unitary operator to describe
the two-qubit gate
is given by\cite{Falci,Zhu_prl2002}
\begin{equation}
\label{two_U1}
U^{tq}=diag(U_{(\gamma^0,\chi^0)},U_{(\gamma^1,\chi^1)}),
\end{equation}
under the condition
that the control qubit is far away from
the resonance condition for 
the operation of the target qubit.
Here $\gamma^{\delta}$ $(\chi^\delta)$ represents the total
phase (the cyclic initial state )
of the target qubit when the
control qubit is in state $\delta(=0,1)$.
Following Ref.\cite{Lloyd}, we find that
unitary operator (\ref{two_U1}) is a nontrivial two-qubit gate
if and only if $\gamma^1 \not= \gamma^0$ or $\chi^1 \not= \chi^0$
$(mod\ 2\pi)$. For example,
\begin{equation}
\label{two_U0}
U^{tq}_{(\gamma^0,\gamma^1)}=
diag(e^{i\gamma^0},e^{-i\gamma^0},e^{i\gamma^1}, e^{-i\gamma^1}),
\end{equation}
when $\chi^1=\chi^0=0$; this gate was proposed to be achieved in the
adiabatic case in the charge qubit\cite{Falci}.
Combining gate (\ref{two_U0}) with single-qubit
operations we are able to perform a gate described by
\begin{eqnarray}
\nonumber
U_{CN}&=&
\left [
I\otimes U^{sq}(\pi/4,\pi/2) \right ]
U^{tq}_{(0,\pi/2)}
\left [ I\otimes U^{sq}(\pi/4,\pi/2) \right ]^\dagger \\
\label{U_CN}
&=& diag(I,i\sigma_x),
\end{eqnarray}
with $I$ as a
$2\times 2$ unit matrix. This gate
is equivalent to the controlled-NOT (CNOT,
which is defined as
$|m\rangle|n\rangle \rightarrow |m\rangle|m\oplus n\rangle$,
where $\oplus$ denotes the addition modulo $2$.) gate
up to an overall phase factor for
the target qubit.
On the other hand,
$U^{tq}_{(0,\gamma^1)}$
become the controlled-phase (CPHASE,
which is defined as
$|m\rangle |n\rangle \rightarrow |m\rangle \exp(imn\phi) |n\rangle$.)
gate by removing a overall phase for the target qubit.

An alternative practical method  to 
achieve the controlled two-qubit gate
is also available under certain conditions. Denoting the
Hamiltonian of the target qubit as $H_t$, we may
produce $H_t=0$ by choosing certain
parameters of $H_t$ when the controlled qubit is in the state
$|0\rangle$, while $H_t$ is able to 
realize a required gate when the controlled qubit is in state
$|1\rangle$. Then the gate in this case is given by 
\begin{equation}
\label{two_U2}
U^{tq}=diag(I,U(\gamma,\chi)),
\end{equation}
where $\gamma$ is the total phase accumulated
in the evolution when the controlled qubit
is in state $|1\rangle$.
Gate (\ref{two_U2}) corresponds to
gate (\ref{two_U1}) for $\gamma^0=0$ and
$\chi_0=\pi/2$.
$U_{CN}$ in Eq.(\ref{U_CN})
may be directly derived when
$\gamma=\chi=\pi/2$.

So far, we have demonstrated that all  elements of
quantum computation may be achievable by using
a pair of orthogonal cyclic states.

\subsubsection{Geometric quantum gates}

The quantum gates $U$ described in Eqs.
(\ref{single_U}),
(\ref{two_U1}) [or (\ref{two_U2})] may be divided into two
categories:
one is referred to as a geometric gate
if the phase in $U$ is a pure geometric evolution operator
($\gamma_d=0$)\cite{G_gate},
and the other is referred to as a dynamic gate 
as long as there exists a nonzero phase
induced from dynamic origin
({\sl i.e.}, $\gamma_d\neq0$).
Geometric quantum computation demands that
 logical gates in computing are
realized by using  geometric phase shifts, so that
it may have the built-in fault-tolerant
advantage due to the fact
that the geometric phases depend only on some
global geometric features.

A key point in geometric quantum computation is to
remove the dynamic phase. We here address two
methods\cite{Falci,Jones,Duan,Wang,Zhu_prl2002}.
A simpler and also practical one is
to  choose some specific external parameters such that
the dynamic phases of
the pair of cyclic states
accumulated in the whole evolution
may be eliminated.
Interestingly, with this method
the corresponding cyclic states in NMR systems
are dark states (the eigenstate with the zero energy eigenvalue),
and thus no dynamic phase is involved.
The dark state method was proposed for geometric quantum
computation with
trapped ions\cite{Duan},
and then described in NMR systems\cite{Wang}.
The other is referred to as a two-loop method:
let the evolution be dragged
by $\hat{H}$ along two closed loops,
with one being in $t\in [0,\tau]$ and
the other  in $t\in (\tau,\tau+\tau^\prime]$.
The dynamic phases accumulated in 
the two loops may be canceled,
while the AA phases will add.

\section{Application to NMR systems}

So far, we have proposed a general scheme to achieve
a universal set of quantum gates based on a pair of
orthogonal cyclic states. It is important to further consider
implementing this scheme with real physical systems. Here, we
illustrate this implementation
using NMR systems\cite{Gershenfeld,Jones}.
Nevertheless, it is worth pointing out that, in principle,
the above theory may be applicable to
other systems which are potentially viable quantum
computer models.

For NMR systems, the magnetic field in Eq.(\ref{H_two_bit})
or (\ref{H_sq}) in a
rotating magnetic field is
given by
\begin{equation}
\label{B_rotating}
{\bf B}(t)=(B_0 \cos\omega t,B_0 \sin\omega t,B_1),
\end{equation}
where $B_{0,1}$ and $\omega$ are constants.
The qubit state $|\psi (t)\rangle$
is described by the Schr\"{o}dinger equation
\begin{equation}
\label{Schrodinger}
i\hbar \frac{\partial }{\partial t}|\psi (t)\rangle=H|\psi (t)\rangle,
\end{equation}
where the Hamiltonian for a single qubit is given by
\begin{equation}
\label{rotated-field}
H=\frac{1}{2}(\omega_0\sigma_x \cos\omega t+\omega_0 \sigma_y
\sin\omega t+\omega_1 \sigma_z)
\end{equation}
with $\omega_i=-g\mu B_i/\hbar$ $(i=0,1)$ and $g$ being
the gyromagnetic ratio.
The Schr\"{o}dinger
equation with Hamiltonian (\ref{rotated-field}) can be solved
analytically\cite{Zhu_prl2000,Rabi}. In terms of explicit form of the 
solution ${\bf n}(\chi,\omega t)$ represented in Ref.\cite{Zhu_prl2000},
it is found that a pair of orthogonal initial states $|\psi_{\pm}\rangle$
with $\chi=\arctan[\omega_0/(\omega_1-\omega)]$ take cyclic
evolutions with the period $\tau=2\pi/\omega$\cite{Wang},
and the evolution
paths are the curves on a Bloch sphere swept by unit vectors
$\pm{\bf n}(\chi,\omega t)$.
Therefore, we may use this pair of $|\psi_{\pm}\rangle$
to achieve single qubit gates described in Eq.(\ref{single_U}),
where the corresponding phases for one cycle are
given by 
\begin{eqnarray}
\label{g_phase_NMR1}
\gamma_g &=& -\pi(1-\frac{\omega_1-\omega}{\Omega}), \\
\label{d_phase_NMR1}
\gamma_d &=& -\pi\frac{\omega_0^2+\omega_1(\omega_1-\omega)}{\omega\Omega},\\
\label{t_phase_NMR1}
\gamma   &=&  -\pi(1+\Omega/\omega), 
\end{eqnarray}
with $\Omega=\sqrt{\omega_0^2+(\omega_1-\omega)^2}$.
In the derivation of the dynamic phase, 
$E_{+}(t)=[\omega_1 \cos\chi+\omega_0 \sin\chi]\hbar/2$
is used.
We may choose any two processes with different
values $\{\omega_0,\omega_1,\omega \}$
satisfying 
Eq.(\ref{criterion})
to accomplish                
two noncommutable qubit gates.

A similar
method may be employed to achieve the two-qubit operation. The
spin-spin interaction in NMR is very well approximated
by
$$
H_I=J\sigma_z^1\sigma_z^2/2.
$$
The state of control qubit is (almost) not
affected by any operation of the target qubit
if  $\omega_1^t$ of the target qubit is chosen to be
significantly different from $\omega_1^c$ of the control qubit.
We may prove that
the initial states
$|\psi_{\pm}\rangle$ described by
$\chi^{\delta}=\arctan [\omega_0/(\omega_1^\delta-\omega)]$
are a pair of orthogonal cyclic states, and may be used
to achieve a two-qubit gate described by
Eq.(\ref{two_U1}).
Here $\omega_1^\delta=\omega_1+(2\delta-1)J$,
$\omega$, $\omega_0$, and $\omega_1$ are parameters for
the target qubit (the superscript "t" is omitted for brevity).
The corresponding phases for one cycle are
given by 
\begin{eqnarray}
\label{g_phase_NMR2}
\gamma^\delta_g &=& -\pi(1-\frac{\omega_1^\delta-\omega}
{\Omega^\delta}),\\
\label{d_phase_NMR2}
\gamma_d^\delta &=& -\pi\frac{\omega_0^2+\omega_1^\delta
(\omega_1^\delta-\omega)}{\omega\Omega^\delta},\\
\label{t_phase_NMR2}
\gamma^\delta   &=& -\pi(1+\Omega^\delta/\omega),
\end{eqnarray}
with $\Omega^\delta=\sqrt{\omega_0^2+(\omega_1^\delta-\omega)^2}$.
It is seen from Eq.(\ref{t_phase_NMR2}) that
the gate described by Eq.(\ref{two_U2})
may be accomplished by choosing the following special parameters,
$$
\omega=\omega_0=\omega_1-J.
$$

It is worth pointing out that
we may achieve the nonadiabatic geometric gates
by choosing some specific parameters, with
which $\gamma_d=0$
in the whole process. 
It is direct to verify that the dynamic phase
in Eq.(\ref{d_phase_NMR1})
is zero
under the following condition,
\begin{equation}
\label{zero_d1_nmr}
\omega=\frac{\omega_0^2+\omega_1^2}{\omega_1}.
\end{equation}
Thus the single qubit gates with the parameters
satisfying Eq.(\ref{zero_d1_nmr})
are geometric quantum gates with
geometric phase shift
$\gamma_g=-\pi (1+\omega_0/\sqrt{\omega_0^2+\omega_1^2})$.
The geometric phases $\gamma_g$ versus the ratio
$\omega_1/\omega_0$ are plotted in Fig. 1.
It is seen that the nontrivial phases 
required for two universal single-qubit gates
may be simply realized by any two processes
with different values of $\omega_1/\omega_0$
(except for zero or infinite).

Besides, the geometric two-qubit gates
are realized whenever
\begin{eqnarray}
\label{zero_d2_nmr}
\omega &=& 2\omega_1,\\
\label{zero_d3_nmr}
\omega_1^2 &=& \omega_0^2+J^2.
\end{eqnarray}
Correspondingly, the conditional geometric phases are
given by $\gamma_g^\delta =-\pi (1+\sqrt{\omega_1^\delta/2\omega)}$.
Figure 2 shows the conditional phases $\gamma_g^\delta$ versus the ratio
$\omega_1/J$.
It is evident that
the nontrivial phases ($\gamma_g^0 \neq \gamma_g^1$ in mod $2\pi$)
required for two-qubit gates
may be achieved for $0<\omega_1/J<\infty$.

It is worth pointing out that the constraint
described by Eq.(\ref{zero_d1_nmr}) [ or Eqs.(\ref{zero_d2_nmr})
and (\ref{zero_d3_nmr})] 
is equivalent to the condition
that the instantaneous
dynamic phase for the wave function in single qubit
(or the target qubit) is always zero\cite{Wang},
namely, the states $|\psi_{\pm}\rangle$ used here 
are the dark states.

The advantage of the above
nonadiabatic gates is that $\omega$ is of the same order
of magnitude as $\omega_0$ or $\omega_1$.
This implies that the speed of geometric quantum gate
is comparable with that of
the dynamic quantum gate. In contrast, the speed of
quantum gate based on
adiabatic Berry's phase is much lower than that of gate
using dynamic phase, since the adiabatic condition 
requires that both $\omega_0$ and $\omega_1$ should be much
larger than $\omega$.

Note that,
the geometric gates $U^{sq}_{1,2}$ may not be
practical by directly using the field ${\bf B}$ in Eq.(\ref{B_rotating})
as the corresponding geometric phase in Eq.(\ref{g_phase_NMR1})
is determined by the angle $\chi$. For example,
Eq.(\ref{g_phase_NMR1})
can be rewritten as $\gamma_g=-\pi (1-\cos\chi)$; thus
$\gamma_g=0$ $(-\pi)$ as $\chi=0$ ($\pi/2$). This problem can
be solved by rotating  the field. It will be seen below that the parameter
$\chi$ for the initial cyclic state may be changed by  rotating
the symmetric axis of field (\ref{B_rotating}), while
the phases in
Eqs.(\ref{g_phase_NMR1}),(\ref{d_phase_NMR1}), and (\ref{t_phase_NMR1})
are invariant.

We introduce a rotation operator ${\bf R}(\hat{y},\chi'-\chi)$
that represents the rotation of  angle $\chi'-\chi$
around the $\hat{y}$-axis, that is,
\begin{equation}
\label{rotation_su2}
{\bf R}^{(2)}(\hat{y},\chi'-\chi)=\exp[-i(\chi'-\chi)\sigma_y]
\end{equation}
in the $SU(2)$ representation, and
\begin{equation}
\label{rotation_so3}
{\bf R}^{(3)}(\hat{y},\chi'-\chi)=\exp[-i(\chi'-\chi)\tau_2]
\end{equation}
with
\begin{equation}
\label{tau_2}
\tau_2=\left (
\begin{array}{llll}
0 & 0 & i  \\
0 & 0 & 0  \\
-i & 0 & 0  \\
\end{array}
\right )
\end{equation}
in the $SO(3)$ representation.
Assuming the required angle
is $\chi'$ in Eq. (\ref{single_U}),
we may apply a magnetic field
${\bf B}^\prime=
{\bf R}^{(3)}(
\hat{y},\chi^\prime-\chi) {\bf B}$,
then the solution of
the cyclic states are
${\bf n}^\prime_{\pm}={\bf
R}^{(3)}(\hat{y},\chi^\prime-\chi)[\pm {\bf n}_{\pm} (\chi,\omega t)]$
[$|\psi_{\pm}\rangle
={\bf R}^{(2)}(\hat{y},\chi^\prime-\chi)|\psi_\pm\rangle$]
because of the spherical symmetry of the system.
Thus $\chi$ may change to
any  required $\chi^\prime$ for  implementation of the quantum
gate, with the geometric phase being unchanged, because the area
swept by ${\bf n}^\prime$ is the same as that by ${\bf n}$.
On the other hand, we have
\begin{eqnarray*}
E^{\prime}_{\pm}(t)
&=& -\mu{\bf n}^\prime_{\pm}(t)\cdot {\bf B}^\prime(t)\\
&=& -\mu [{\bf R}^{(3)}(\hat{y},\chi^\prime-\chi)
{\bf n}_{\pm}(t)]\cdot
[{\bf R}^{(3)}(\hat{y},\chi^\prime-\chi)
{\bf B}(t)]\\
&=& E_{\pm}(t).
\end{eqnarray*}
Therefore, we have proven that the invariant
of all phases with respect to the rotation of the symmetric axis of
the field in Eq. (\ref{B_rotating}).
We conclude that
$\gamma$ $(\chi)$ in gate
(\ref{single_U}) is determined by the
values of $\{ \omega_0,\omega_1,\omega \}$
(the symmetric axis of the magnetic field).
For example, if the magnetic field is ${\bf B}^\prime$ for
$\chi^\prime=0$ ($\pi/2$), we may achieve the geometric gate
$U_1$ ($U_2$) with $\gamma_{1,2}=-\pi(1-\cos\chi)$.
It is worth pointing out that
the above method to control 
$\chi$ and $\gamma_g$ separately in quantum gates
is also feasible in nongeometric gates.

We turn to address how to remove the dynamic phases
in a multiloop nonadiabatic evolution.
The possible generalization of a multiloop method
from the adiabatic evolution\cite{Falci,Jones}
to nonadiabatic case was mentioned in Refs.\cite{Wang,Zhu_prl2002}.
We here wish to demonstrate explicitly
one removal procedure of the dynamic phase.

Let us first choose
the magnetic fields in two loops as
\begin{eqnarray}
\label{loop1}
Loop\ 1:\ &B&=(B_0 \cos\omega t,B_0 \sin\omega t , B_1),
\ t\in [ 0, \tau) \\
\nonumber
Loop\ 2:\ &B&^\prime={\bf R}^{(3)}(\hat{y},\alpha^\prime-\alpha)
(-B_0^\prime \cos\omega t, -B_0^\prime \sin\omega t, -B_1^\prime),\\
\ &t&\in [\tau,2\tau]
\label{loop2}
\end{eqnarray}
where
$\tau=2\pi/\omega$,
$\alpha=\arctan[\omega_0/(\omega_1-\omega)]$,
and $\alpha\prime=\arctan[\omega_0^\prime/(\omega_1^\prime+\omega)]$
with $\omega_i^\prime=-g\mu B_i^\prime/\hbar$ $(i=0,1)$.
As shown before, a pair of orthogonal initial
states $|\psi_{\pm}\rangle$ ($|\psi_{\pm}^\prime\rangle$) with
$\chi=\alpha$ ($\chi^\prime=\alpha^\prime$) take cyclic evolutions
during the loop one (two).
The rotation
${\bf R}^{(3)}(\hat{y},\alpha^\prime-\alpha)$ in Eq.(\ref{loop2})
ensures that
the cyclic initial states in the two loops are
the same at the time $t=2\pi/\omega$ \cite{Rotation}. Therefore,
the gate described by the two loops is given by
$U=U(\chi,\gamma^{(1)}+\gamma^{(2)})$, where
$\gamma^{(1)}$ ($\gamma^{(2)}$) is the total phase
accumulated in the loop one (two). Denoting
$\gamma_d^{(l)}$ $(l=1,2)$ and $\gamma_g^{(l)}$
the dynamic phases and geometric phases
accumulated in the loop $l$, respectively,
we now illustrate that there exist
processes satisfying
\begin{eqnarray*}
\gamma_g^{(1)}+\gamma_g^{(2)}&=&-\Gamma\pi,\\
\gamma_d^{(1)}+\gamma_d^{(2)}&=&0,
\end{eqnarray*}
where $-\Gamma\pi$ is a nontrivial geometric phase which
we intend to realize in geometric quantum gates.
Then the magnetic fields should
satisfy the following equations
\begin{eqnarray}
\label{remove1}
\frac{\omega_1-\omega}{\Omega}
&+&
\frac{\omega_1^\prime+\omega}
{\Omega^\prime}=2-\Gamma,\\
\label{remove2}
\frac{\omega_0^2+\omega_1^2-\omega\omega_1}{\omega\Omega}
&=&
\frac{(\omega_0^\prime)^2+(\omega_1^\prime)^2+\omega\omega_1^\prime}
{\omega\Omega^\prime},
\end{eqnarray}
where $\Omega=\sqrt{\omega_0^2+(\omega_1-\omega)^2}$ and
$\Omega^\prime=\sqrt{(\omega_0^\prime)^2+(\omega_1^\prime+\omega)^2}$.
As for the required $\Gamma$,
it is possible that
there exist many
solutions, since
there are
five unknown variables in two equations.
For  example,
we numerically
calculate the solutions
for $\Gamma=1/2$.
For simplicity,  we set $\omega_1^\prime=\omega_1$ as the unit,
and find that if $\{\omega,\omega_0,\omega_0^\prime\}$
$(0< \omega < \omega_1)$
satisfy the equations given by
\begin{eqnarray*}
\omega &+& 1.13389\omega_0=0.99998,\\
\omega &+& 1.07091\omega_0-0.06299\omega_0^\prime=0.88889,
\end{eqnarray*}
which describe a straight line(segment) in the
three-dimensional space, the geometric phase accumulated in
the whole two-loop evolution is just what we required, with
the total dynamic phase being zero.

The multiloop method to remove dynamic phase is also
feasible for two-qubit geometric quantum gates.
We choose the magnetic fields on the target qubit
in two loops as
\begin{eqnarray}
\label{loop2q_1}
Loop\ 1:\ &B&=(B_0 \cos\omega t,B_0 \sin\omega t , B_1),
\ t\in [ 0, \tau), \\
\nonumber
Loop\ 2:\ &B&^\prime={\bf R}^{(3)}(\hat{y},\eta)
(-B_0^\prime \cos\omega^\prime t, -B_0^\prime \sin\omega^\prime t, -B_1^\prime),\\
\ &t&\in [\tau,\tau+\tau^\prime],
\label{loop2q_2}
\end{eqnarray}
where
$\tau^\prime=2\pi/\omega^\prime$.
The angle
\begin{equation}
\label{Eta}
\eta=\arctan[\omega_0/(\omega_1^\delta-\omega)]
-\arctan[\omega_0^\prime/(\omega_1^{\prime\delta}+\omega^\prime)]
\end{equation}
should be independent on the state $\delta$ in the control qubit.
To guarantee that the interaction between qubits
is still determined by the original initial state
$\delta$ of the control qubit,
the control qubit should be rotated by
${\bf R}^{(3)}(\hat{y},\eta)$ at time $t=\tau$
[The state of the controlled qubit is unchanged if
a rotation ${\bf R}^{(3)}(\hat{y},-\eta)$ is also
applied at time $\tau+\tau^\prime$].
Correspondingly, the $\delta-$independent constraint
described by Eq.(\ref{Eta}) can be rewritten as
\begin{equation}
\label{Constrain}
\frac{\omega_0}{(\omega_1-\omega)^2-J^2}=
\frac{\omega_0^\prime}{(\omega_1^\prime+\omega^\prime)^2-J^2}.
\end{equation}
On the other hand, the condition under which there exist processes
with zero dynamic phase is
\begin{equation}
\label{Constrain1}
\frac{\omega_0^2+\omega_1^\delta(\omega_1^\delta-\omega)}
{\omega\Omega^\delta}=
\frac{(\omega_0^\prime)^2+\omega_1^{\prime\delta}
(\omega_1^{\prime\delta}+\omega^\prime)}
{\omega^\prime\Omega^{\prime\delta}},
\end{equation}
where $\Omega^\delta=\sqrt{\omega_0^2+(\omega_1^\delta-\omega)^2}$ and
$\Omega^{\prime\delta}=\sqrt{(\omega_0^\prime)^2+(\omega_1^{\prime\delta}
+\omega^\prime)^2}$. Note that
the geometric phases are nontrivial ( $\gamma_g^1\neq \gamma_g^0$ in
$mod\  2\pi$), and thus can be
applicable in geometric quantum computation.

The magnetic fields, which satisfy Eqs.
(\ref{Constrain}) and (\ref{Constrain1}) in loop two,
as a function
of $\omega$ are plotted in Fig.3a, where
$\omega_0=\omega_1=5.0$ with
$J$ as the unit.
We may numerically calculate the three unknown variables
$\{ \omega^\prime, \omega_0^\prime,\omega_1^\prime \}$
in three equations described by
Eqs. (\ref{Constrain}) and (\ref{Constrain1}).
Then the conditional geometric phases may be obtained from
equations $\gamma_g^\delta=-\Gamma^\delta\pi$ with
$$
\Gamma^\delta=2-\frac{\omega_1^\delta-\omega}{\Omega^\delta}
-
\frac{\omega_1^{\prime\delta}+\omega^\prime}
{\Omega^{\prime\delta}}.
$$
The corresponding conditional phases $\gamma_g^\delta$
for $\omega_0=\omega_1=5.0$ as a function of $\omega$
are plotted
in Fig.3(b). It is seen that the nontrivial phases
$\gamma^1_g\neq \gamma_g^0$ may be realized
by appropriately choosing the values of 
$\{ \omega, \omega_0,\omega_1 \}$ and
$\{ \omega^\prime, \omega_0^\prime,\omega_1^\prime \}$.
As a consequence the nontrivial two-qubit geometric quantum
gate may be achieved.

\section{Conclusions and discussions}

An experimentally feasible scheme 
based on a pair of orthogonal cyclic states
has been proposed to
accomplish a universal set of quantum logical gates,
in which quantum computation implemented by
both dynamic and geometric operations can work,
i.e., quantum gates in this scheme
may be implemented by
the total phases accumulated in the cyclic evolution, and
the geometric quantum computation can be achieved by eliminating
the dynamic phase.
Furthermore, the geometric phase shift used is 
the cyclic AA phase, which can be
nonadiabatic.
It is possible that the gates achieved here
can handle arbitrary quantum computation
without the intrinsic limitation on operation time.
Therefore, the nonadiabatic method proposed here may allow us to
physically implement (geometric) quantum computation even for
systems with very short decoherence time, which could be
especially useful for solid-state implementations of scalable
quantum computers.

We here discuss briefly the errors induced by random noises
in geometric quantum computation.
Random noises may lead to two kinds of errors.
One is that the path
may not be exactly closed at the end of the gate operation,
leading to the noncyclic corrections. The other is
that the evolution path may fluctuate around the ideal path
with known cone angle.
The noncyclic corrections could be
negligible at least when the first-order corrections from
random noises are taken into account,
as was indicated in Ref. \cite{Palma}.
On the other hand,
as that in the adiabatic cyclic geometric scheme,
the present scheme is also robust to the second type of errors as
the area enclosed by the evolution path
(geometric phase) is insensitive to
the random fluctuation.

Finally, we wish to make a few remarks on
experimental implementation of
geometric quantum computation. 
The simplest geometric quantum computation
should experimentally complete
the following three steps one by one:
(i) detection of the (conditional) geometric phase shifts in
qubit systems;
(ii) implementation of a universal set of geometric quantum logic
gates, particularly the implementation of a conditional gate. 
(iii) illustration of a simple algorithm by pure geometric quantum gates, 
such as Deutsch's problem, Grover's search algorithm, or
Shor's factorization algorithm, etc..
Two recent exciting
experiments reported that
the conditional geometric
phase shifts for quantum logical gates
using NMR were detected
in adiabatic\cite{Jones} and nonadiabatic\cite{Du} regions.
However, a universal set of gates as well as
a simple quantum algorithm experimented  
by (adiabatic or nonadiabatic) geometric phases
are still awaited.

\acknowledgements{
We are very grateful to Dr. L. M. Duan, Dr. J. W. Pan
and Dr. X. B. Wang for
valuable discussions,
and Dr. P. Zanardi for pointing
out Ref.\cite{Du}.
This work was supported by the RGC grant of Hong Kong under
Grants Nos. HKU7118/00P and HKU7114/02P,
 and a CRCG grant at the HKU.  
S. L. Z. was supported in part by SRF
for ROCS, SEM, the NSF of
Guangdong Province under Grant No. 021088,
and the NNSF of China under Grant No. 10204008.
}

\begin{figure}[tbp]
\label{fig1}
\vspace{-0.6cm}
\epsfxsize=7.5cm
\epsfbox{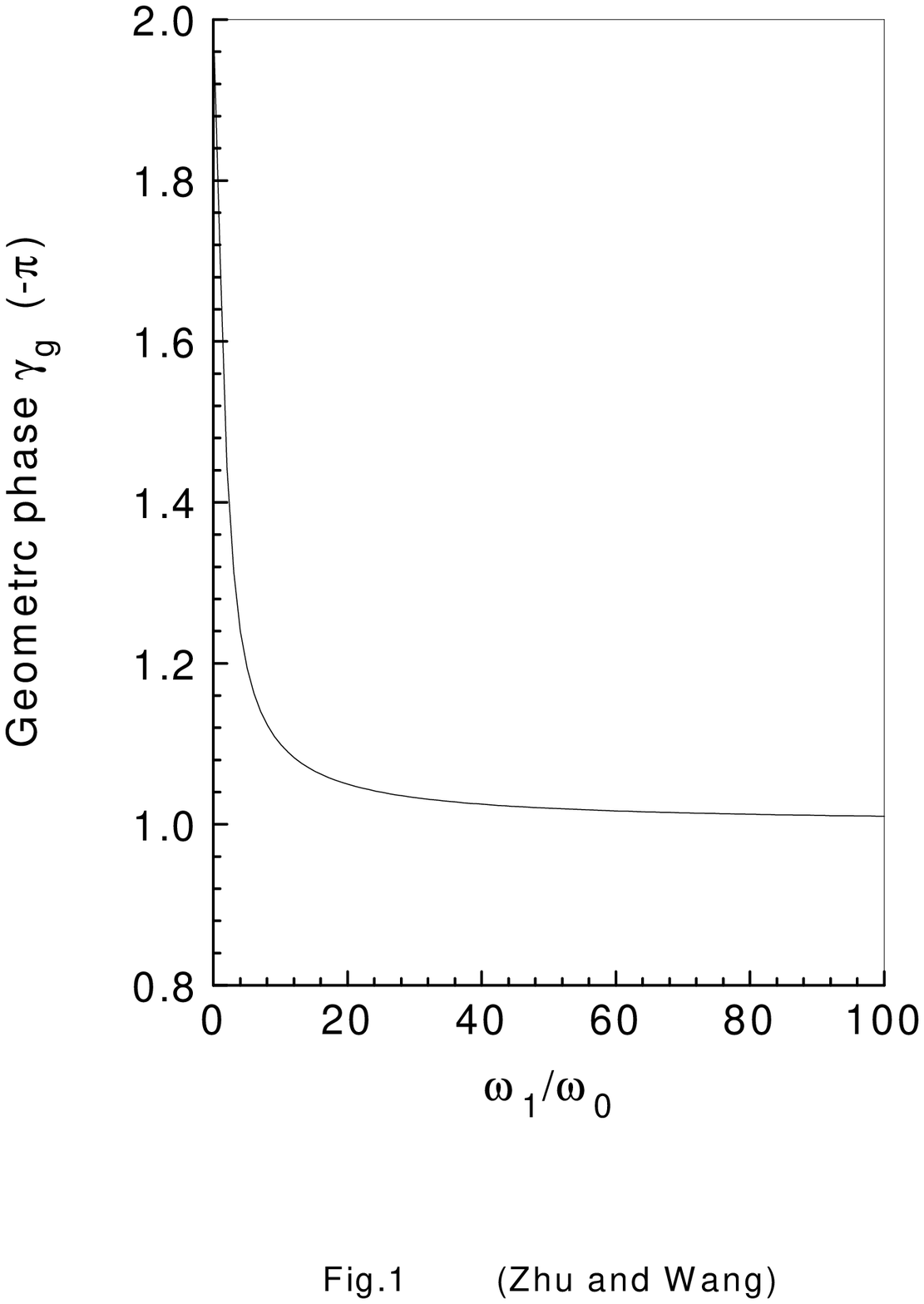}
\vspace{-0.3cm}
\caption{The geometric phase $\gamma_g$ versus the ratio
$\omega_1/\omega_0$.
}
\end{figure}

\begin{figure}[tbp]
\label{fig2}
\vspace{-0.6cm}
\epsfxsize=7.5cm
\epsfbox{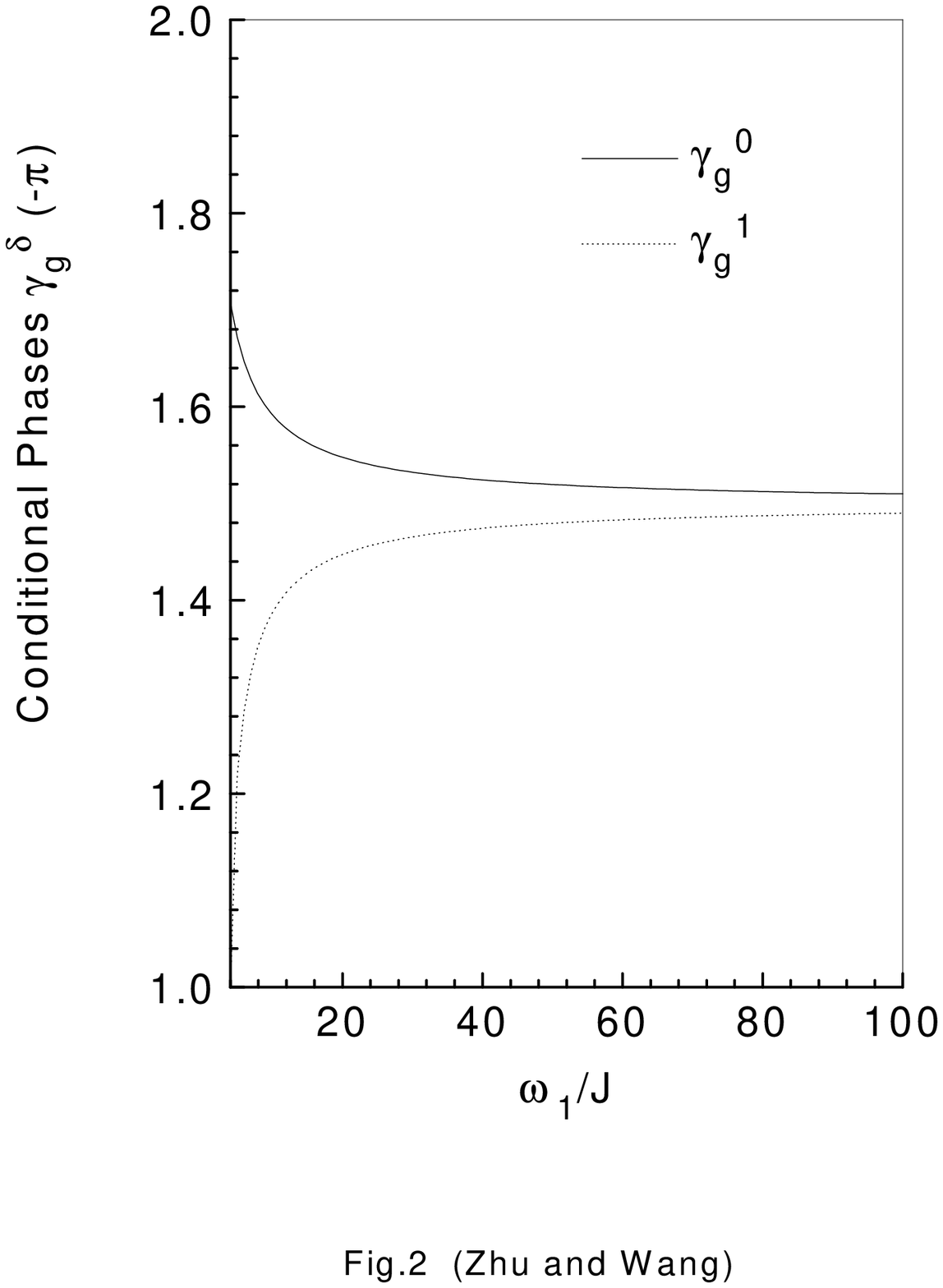}
\vspace{-0.3cm}
\caption{The conditional phases $\gamma_g^\delta$ versus the ratio
$\omega_1/J$.
}
\end{figure}

\begin{figure}[tbp]
\label{fig3a}
\vspace{-0.6cm}
\epsfxsize=7.5cm
\epsfbox{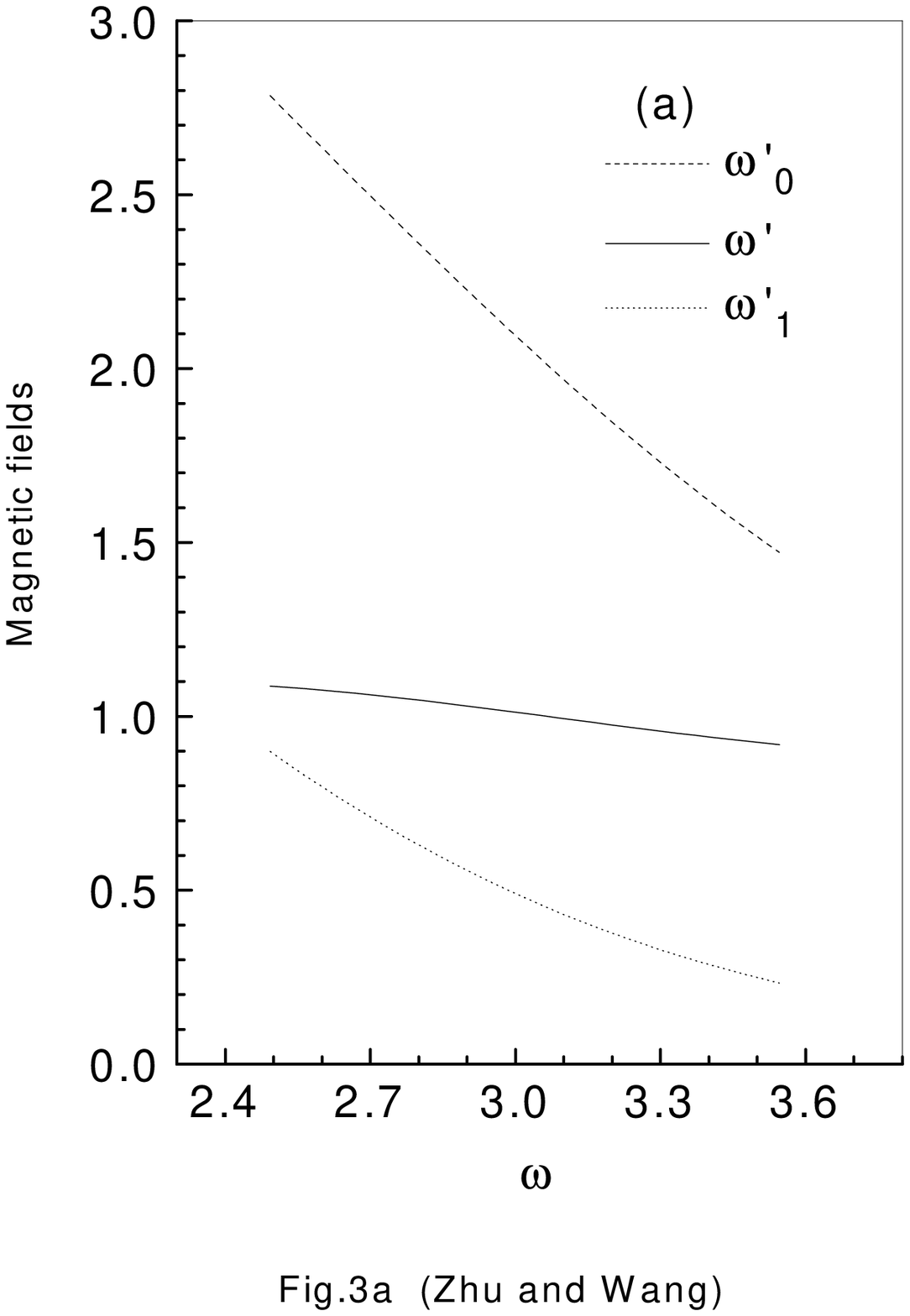}
\vspace{-1.0cm}
\epsfxsize=7.5cm
\epsfbox{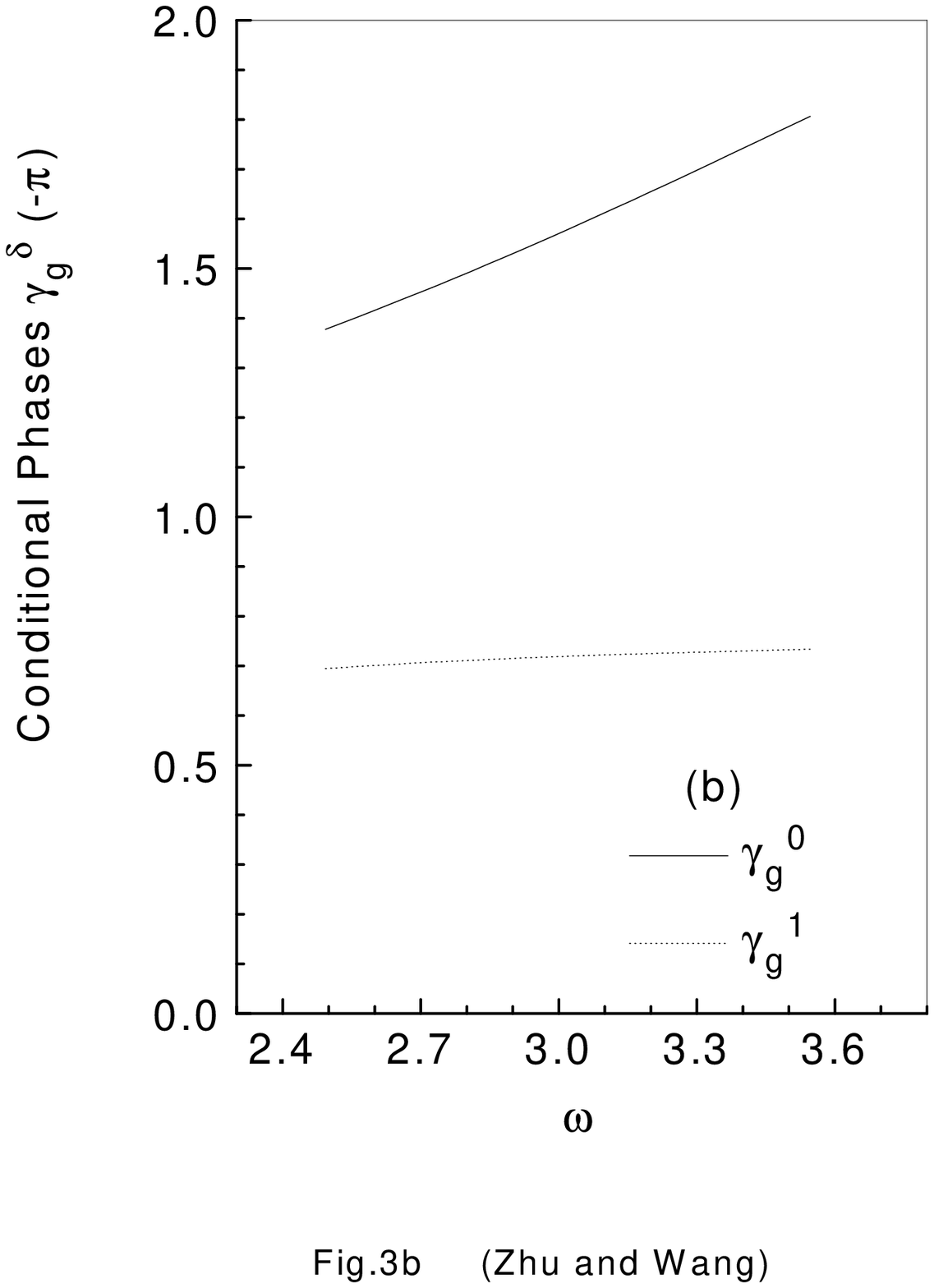}
\caption{ (a) The magnetic fields required in loop two
versus $\omega$. (b)
The conditional geometric phases $\gamma_g^\delta$ versus $\omega$.
}
\end{figure}


\begin{references}
\bibitem{G_gate} Here the dynamic phase accumulated in
the gate operation is possibly nonzero when an input state is
a superposition of $|\psi_{\pm}\rangle$.
However, it is possible to define a more
strictly geometric gate $U$ where the phase accumulated
is a pure geometric phase for
any input state. The latter gates have not been studied here.
\end{references}
\end{document}